\documentstyle[12pt]{article}
\def\be{\begin{eqnarray}}
\def\ee{\end{eqnarray}}
\def\ba{\begin{array}}
\def\ea{\end{array}}

\begin{document}

\begin{center}
{\bf\Large {Generalized Quantum Dynamics with Arrow of Time}}
\end{center}

\vskip 1cm

\begin{center}
{\bf \large {Vadim V. Asadov$^{\star}$\footnote{asadov@neurok.ru}
and Oleg V.
Kechkin$^{+\,\star}$}\footnote{kechkin@depni.sinp.msu.ru}}
\end{center}

\vskip 5mm

\begin{center}
$^+$Institute of Nuclear Physics,\\
Lomonosov Moscow State University, \\
Vorob'jovy Gory, 119899 Moscow, Russia
\end{center}

\vskip 2mm

\begin{center}
$^\star$Neur\,OK--III,\\
Scientific park of MSU, Center for Informational Technologies--104,\\
Vorob'jovy Gory, 119899 Moscow, Russia
\end{center}

\vskip 1cm

\begin{abstract}
It is shown, that quantum theory with complex evolutionary time
parameter and non-Hermitian Hamiltonian structure can be used for
natural unification of quantum and thermodynamic principles. The
theory is postulated as analytical in respect to the parameter of
evolution, which real part is identified with the `usual'\, physical
time, whereas the imaginary one is understood as proportional to the
inverse absolute temperature. Also, the Hermitian part of the
Hamiltonian is put equal to conventional operator of energy. It is
shown, that the anti-Hermitian Hamiltonian part, which is taken as
commuting with the energy operator, is constructed from parameters
of decay of the system. It is established, that quantum dynamics,
predicted by this theory, is integrable in the same sense as the
corresponding non-modified one, and that it possesses a well defined
arrow of time in isothermal and adiabatic regimes of the evolution.
It is proved, that average value of the decay operator decreases
monotonously (as the function of the physical time) in these
important thermodynamical regimes for the arbitrary initial data
taken. We discuss possible application of the general formalism
developed to construction of time-irreversible modification of a
string theory.
\end{abstract}

\vskip 0.5cm

PACS No(s).\, : 05.30.-d,\,\,05.70.-a.

\renewcommand{\theequation}{\thesection.\arabic{equation}}


\section{Introduction}

Irreversibility of evolution is a common feature of all real
dynamical systems. This fact is reflected in the second law of
thermodynamics, which states that entropy of any closed physical
system can not decrease. Figuratively, one says about `arrow of
time',\, which separates past and future in absolute manner, and
must be guaranteed by any realistic dynamical theory without fail
\cite{atf}--\cite{atl}.

It is important to stress, that thermodynamics itself is not a
dynamical theory of the same type as, for example, classical or
quantum mechanics. Rather, it provides general theoretical framework
for dynamical theories, which pretend to adequate description of
evolving physical reality. Also, it can be said, that
thermodynamical principles must be realized on the base of
consistent dynamical theory of fundamental type, and that the
thermodynamics imposes hard restrictions to the corresponding
theoretical constructions \cite{td}.

In this connection, one must take into account, that all quantum and
classical fundamental theories are conservative Hamiltonian systems
\cite{qm}--\cite{cm}. All these systems are reversible in time,
because one can interchange their initial and final data to obtain
really possible result of the dynamical evolution. Thus, these
theories, being the most natural concrete classes of the closed
dynamical systems, can not be used for any consistent realization of
thermodynamic conception of the arrow of time. All attempts
performed to `average out'\, the reversible results of these
theories to obtain the thermodynamic irreversible ones, contain
hidden incorrect actions. For example, impossibility to consider
unidirectional evolution in framework of classical mechanics can be
understood using results of the Poincare theorem. Actually, it
states, that one can decompose any classical motion to the set of
`Poincare cycles',\, so this motion obtains transparently reversible
character. The same situation takes place in the quantum mechanics:
using the Hamiltonian proper basis, one reduces the quantum motion
to the set of corresponding oscillations, which are reversible in
time manifestly. Also, probabilities to find quantum system in the
Hamiltonian eigenstates remain constant quantities, and one can not
relate any irreversible dynamics of the kinetic type to them.

It is important to emphasize, that the set of conventional kinetic
equations contain arrow of time by its definition. Such equations
had been used extensively by Prigogine in his study of both
thermodynamic and synergetic processes \cite{Pr}. The only
conceptual problem of such approaches is how to ground the kinetic
method of realization of the time irreversible disciplines on the
fundamental level in the really correct form. In the over words, one
must show how to modify the fundamental theories of the classical
and quantum mechanical types to achieve their consistency with the
conventional kinetic framework. In this paper we answer on this
question by presentation of some `minimal'\, generalization of the
quantum theory. The corresponding modification of the classical
theory will be developed in forthcoming publications.

Our generalization of the conservative quantum theory is related to
use of complex parameter of the evolution (or `complex time',\, see
\cite{ctf}--\cite{ctl}), and non-Hermitian Hamiltonian operator
independent on this parameter \cite{NonHerm-f}--\cite{NonHerm-l}.
Also, we suppose analytic dependence of quantum states of the theory
on the complex time, and impose commutativity restriction on the
Hamiltonian and result of its Hermitian conjugation. We interpret
the real part of the complex time as `usual'\, physical time,
whereas the imaginary one we identify as proportional to the inverse
absolute temperature of the system. Then, the Hermitian part of the
Hamiltonian we put equal to energy operator of this quantum system,
whereas the anti-Hermitian part is naturally related to operator of
decay parameters of the energy eigenstates. Finally, we define
thermodynamic regimes of the evolution: this allows us to determine
temperature functions for concrete thermodynamic processes
considered. It is shown, that quantum dynamics, predicted by the
theory, has a well defined arrow of time in the isothermal and
adiabatic regimes of the evolution.

\section{Generalized quantum theory}

\setcounter{equation}{0}

It seems clear, that any realistic generalization of the quantum
theory must start with some complex linear space of state vectors
$\Psi_{1},\,\,\Psi_{2}$, etc., and some well-defined scalar products
$\Psi_{1}^+\Psi_{2}$, etc., of these vectors. Moreover, one must
preserve a probability interpretation of the theory. For example, in
the normalizable case, one must use the relation \be\label{G1}
P=\frac{\Psi_{1}^+\Psi_{2}\,\cdot\,\Psi_{2}^+\Psi_{1}}{\Psi_{1}^+\Psi_{1}\,
\cdot\,\Psi_{2}^+\Psi_{2}} \ee for the probability $P$ to find the
system in the condition with the state vector $\Psi_{1}$, if this
system is specified by the state vector $\Psi_{2}$. Then, it is
necessary to deal with some set of the observables $Q_1,\,Q_2$,
etc., which are linear Hermitian operators acting on the state
vectors. Without any doubt, a rule for calculation of the average
value $\bar Q$ for the observable $Q$, which is related to the state
vector $\Psi$, must save its well-known conventional form:
\be\label{G2} \bar{Q}=\frac{\Psi^+{{Q}}\Psi}{\Psi^+\Psi}. \ee Also,
the average values must preserve their meaning in relation of
results of the quantum theory to the corresponding observations in
the real world.

Then, all dynamical aspects of generalized quantum theory must be
described in terms of some evolutionary parameter $\tau$ and
Hamiltonian operator ${\cal{H}}$. For natural analogies of
conservative systems, one must put ${\cal{H}}_{,\tau}=0$ in the
Schr\"{o}dinger's picture (which we explore in this work). Also, the
main dynamical equation of the theory must preserve its conventional
Schr\"{o}dinger's form, \be\label{G3} i\hbar
\Psi_{,\tau}={\cal{H}}\Psi. \ee Note, that all elements of the
quantum theory listed above are completely standard ones.

Our modification of this theory is based on the use of the complex
time parameter $\tau\neq\tau^*$ and non-Hermitian Hamiltonian
operator $\cal{H}\neq\cal{H}^+$. We consider a holomorphic variant
of the theory, with $\Psi_{,\tau^{*}}=0$ and
${\cal{H}}_{,\tau^{*}}=0$, which leads to the simplest
generalization of the standard theoretical quantum scheme. We
restrict our consideration by the theories with $\left [
\cal{H},\,\,\cal{H}^+\right ]=0$. The last relation becomes an
identity in the standard theory case, when ${\cal{H}}={\cal{H}}^+$,
so one can really speak about the `minimal generalization'. We
parameterize the complex evolutionary parameter $\tau$ in terms of
the real variables $t$ and $\beta$ in the following form:
\be\label{G5} \tau=t-\,i\,\frac{\hbar}{2}\,\beta, \ee whereas the
non-Hermitian operator ${\cal{H}}$ will be represented using the
Hermitian ones $E$ and $\Gamma$ as \be\label{G6}
{\cal{H}}=E-\,i\,\frac{\hbar}{2}\,\Gamma. \ee Note, that
\be\label{G4'}[E,\,\,\Gamma]=0, \ee in accordance to the restriction
imposed above. In the next section we will argue, that, if one
identifies the quantities $t$ and $E$ with the `usual'\, time and
energy operator, respectively, then the remaining quantities $\beta$
and $\Gamma$ mean the inverse absolute temperature $\beta=1/kT$
(multiplied to the Bolzman constant $k$), and the operator of
inverse decay time parameters of the system. Also, it will be
demonstrated, that the former operator defines an arrow of time,
which seems naturally in view of transparent irreversibility of any
decay process.

Then, this scheme of generalization of the quantum theory must be
completed by introducing of a conception of thermodynamic regime,
which has a form of fixation of the temperature function
$\beta=\beta (t)$. For example, one can study evolution of the
system in the isothermal case $\beta=\rm{const}$. Also, it is
possible to analyze the adiabatic situation with
$\bar{E}(t,\beta)=\rm{const}$ (the system under consideration does
not perform a work, because ${\cal H}$ is time-independent, so this
regime is adiabatic actually). The most general thermodynamic
regime, which is related to the observable $Q$, can be defined as
\be\label{G7} f\left ( t,\, \beta,\,\bar{{Q}}(t,\beta\right )=0. \ee
In this work we would like to show, that the modified quantum theory
presented above possesses a well defined arrow of time in the
isothermal and adiabatic regimes of its thermodynamical evolution.


\section{Time, energy, temperature and parameters of decay}

\setcounter{equation}{0}

So, our main goal is to detect and to study irreversible aspects of
evolution of the generalized quantum system, defined in the previous
section. To do it in explicit form, let us express all significant
quantities of the theory in terms of a common basis of the
eigenvectors $\psi_n$ of the commuting operators $E$ and $\Gamma$.
We take it in orthonormal form (i.e., we mean that the identities
$\psi_n^+\psi_k=\delta_{nk}$ take place). Here, of course, the
indexes are understood in the appropriate multi-index sense (and all
summations have the corresponding type). The eigenvalue problem
under consideration reads: \be\label{S1} E\psi_n=E_n\psi_n, \qquad
\Gamma\psi_n=\Gamma_n\psi_n; \ee it can be reformulated in terms of
the non-Hermitian operator ${\cal H}$. Actually, it is easy to see,
that $\psi_n$ is the eigenvector for this operator, which
corresponds to the complex eigenvalue ${\cal
H}_{n}=E_n-i\hbar/2\,\Gamma_n$. Then, the state vectors
$\Psi_n=e^{-i{\cal H}_{n}\tau/\hbar}\psi_n$ satisfy the
Schr\"{o}dinger's equation (\ref{G3}), and also form the complete
(but $\tau$-dependent) basis. This basis can be used for
representation of any solution $\Psi$ of the Schr\"{o}dinger's
equation in the form of linear combination with some set of constant
parameters $C_n$, i.e., as \be\Psi=\sum_n C_n\Psi_n.\ee Using this
decomposition formula and the orthonormal basis properties, one can
calculate the probability $P_n$ to find the quantum system in its
basis state $\Psi_n$, when it is described by the state vector
$\Psi$. After the application of Eq. (\ref{G1}) one obtains, that
\be\label{S2} P_n=\frac{w_n}{Z}, \ee where $Z=\sum_n w_n$,
\be\label{S3} w_n=\rho_n e^{-(E_n\beta+\Gamma_nt)}, \ee and
$\rho_n=|C_n|^2$. Note, that all following analysis will be related
to study of dynamics of the probabilities (\ref{S2})--(\ref{S3}) in
different thermodynamical regimes and for various special
realizations of the generalized quantum theory. In this analysis,
the formula (\ref{S3}) provides the base for both theoretical and
experimental study of the our generalization approach comparing with
the standard quantum theory.

First of all, let us consider evolution of two very special systems,
which dynamical properties can explain our interpretation of the
imaginary part of the complex time parameter $\tau$, and the
anti-Hermitian part of the Hamiltonian operator $\cal H$. Namely,
our first system has coinciding eigenvalues $\Gamma_n$ for the all
indexes $n$ (i.e., this system is `decay-free',\, in fact). It is
easy to see, that \be w_n=\rho_ne^{-E_n\beta}\ee in this case. This
formula demonstrates, that the quantity $\beta$ is actually the
inverse absolute temperature (multiplied to the Bolzman constant),
if $E_n$ is identified with the $n$-th `energy level'\, of the
system.

The second special system is specified by coinciding eigenvalues
$E_n$ of the energy operator (i.e., dynamics of this system is
defined by the operator $\Gamma$ only). Then,
$w_n=\rho_ne^{-\Gamma_nt}$ in this case, so the quantities
$\Gamma_n$ have the sense of decay parameters, if $t$ means the
conventional (`usual') physical time. Actually, let us consider, for
example, the system with \be\label{nstar}\Gamma_{n_{\star}}=\min_n
\{\Gamma_n\}=0\ee in the situation, where the single level $n$ is
weakly excited under the level $n_{\star}$ (in fact, these
`levels'\, are the corresponding solution subspaces). In this
special case, $\rho_{m}=\rho_{n}\delta_{mn}$, where $n,m \neq
n_{\star}$, and also $\rho_{n_{\star}}\approx 1$, whereas
$\rho_{m}\approx 0$ (because $\rho_{m}<<\rho_{n_{\star}}$). It is
easy to see, that for this weakly excited quantum state, \be
P_{m}\approx\rho_{m}e^{-\Gamma_{m}t},\ee so $t_{m}=1/\Gamma_{m}$ is
a conventional `time of life'\, for the excitation under
consideration, if $t$ has the standard time interpretation. Also, it
is clear, that the subspace marked by the index $n_{\star}$\, plays
an attractor role in the dynamics of the second special system.
Note, that in the same situation with $\Gamma_{n_{\star}}=\max_n
\{\Gamma_n\}$, one deals with the exponentially increasing
probability $P_{m}(t)$. However, we use the term `parameter of
decay'\, for the quantity $\Gamma_n$ in all regimes of the
evolution.

Then, the solution space of the theory of a discussing type can be
decomposed into the direct sum of the subspaces, which have a given
value of the energy or of the parameter of decay. For all these
subspaces, the interpretation of $\beta$ and $\Gamma_n$ is the same
one, as for the special systems of the first and second discussed
types, respectively. Finally, we extend the interpretation of these
physical quantities (as well as the interpretation of the quantities
E and t) to the total solution space of the generalized quantum
theory, making a simple and natural fundamental generalization.

\section{Quantum (thermo)dynamics and arrow of time}

\setcounter{equation}{0}

Now let us consider non-specified quantum theory of the form,
presented in the previous sections, and study the isothermal regime
of its thermodynamical evolution. It is easy to prove, that in the
case of $\beta=\rm{const}$, the dynamical equation for the basis
probabilities reads: \be\label{S4} \frac{dP_n}{dt}=-\left (
\Gamma_n-\bar{\Gamma}\right )P_n .\ee To perform its analysis, let
us study a behavior of the quantity $\bar{\Gamma}$. After some
calculations one obtains, that \be\label{S5}
\frac{d\bar{\Gamma}}{dt}=-D_{\Gamma}^2,\ee where
$D_{\Gamma}^2=\overline {\left ( \Gamma-\bar{\Gamma}\right )^2}$ is
the squared dispersion of the quantum observable $\Gamma$. From Eq.
(\ref{S5}) it follows, that the function $\bar\Gamma (t)$ is not
increasing. This means, that the isothermal regime of evolution has
arrow of time. It is seen, that this can be related to the average
value of the decay operator of this quantum system.

Then, the probability $P_n$ rises, if $\bar{\Gamma}>\Gamma_n$, and
degenerates, when $\bar{\Gamma}<\Gamma_n$. Thus, in the isothermal
case, the quantity $|\Gamma_n-\bar{\Gamma}|$ has a sense of the
inverse time of exponential growth or degeneration of the
probability to find the system in the subspace of ejgenstates with
the given value $\Gamma_n$ of the decay operator $\Gamma$. Also, in
this regime, one obtains the following picture for asymptotics of
the probabilities: all `activated'\, probabilities (with $\rho_n\neq
0$) with the maximal value of the decay parameter rise droningly,
whereas all the remaining probabilities fall to the zero values. The
non-activated probabilities (with $\rho_n= 0$) remain trivial during
all dynamic history of the system.

Actually, let us define the multi-index $n_{_\star}$ by the relation
(\ref{nstar}), again. Then, for the only non-degenerating (at
$t\rightarrow +\infty$) probabilities $P_{n_{\star}}$, one obtains
the following asymptotical result: \be\label{S6}
P_{n_{_\star}}(+\infty)=\Pi P_{n_{_\star}}(0), \ee where the scale
parameter $\Pi>1$ reads: \be\label{S7} \Pi=1+\frac{\sum_{n\neq
n_{_\star}}\rho_ne^{-E_n\beta}}
{\sum_{n_{_\star}}\rho_{n_{_\star}}e^{-E_{n_{_\star}}\beta}}. \ee
Note, that the relations (\ref{S6})--(\ref{S7}) have a form of \,
`dressing procedure'\, in standard quantum field theory. This
circumstance seems really hopeful in context of solution of
different problems related to its renormalization. Roughly speaking,
in such theories one deals with infinite set of oscillating
harmonics (in the corresponding representation on shell). Then, one
needs in cut of this infinity to reach a theoretical scheme with
really consistent calculations. However, all known cut procedures
are in contradiction with `all normal neglecting principles'.\, The
generalized quantum theory presented above allows one to work with
the modes, which degenerate dynamically, and also with the ones,
which remain `alive'\, at the `big times'.\, Moreover, these former
modes of the exact quantum theory solution become renormalized
during the total dynamical history of the system, as it follows from
comparison of values for their initial and final probabilities (see
Eqs. (\ref{S6})--(\ref{S7})).

Then, it is easy to prove, that in the general case of $\beta=\beta
(t)$, the dynamical equations for the basis probabilities have the
following form: \be\label{S8} \frac{dP_n}{dt}=-\left [
\Gamma_n-\bar{\Gamma}+\left ( E_n-\bar{E}\right
)\frac{d\beta}{dt}\right ] P_n. \ee Here, the specific function
$d\beta/dt$ must be extracted from the corresponding thermodynamical
regime (\ref{G7}). In the adiabatic case, when
$\bar{E}=\sum_nE_nP_n=\rm{const}$, one obtains immediately, that
\be\label{S9} \frac{d\beta}{dt}=-\frac{\overline{E\Gamma}-\bar
E\bar\Gamma}{D_{E}^2}, \ee where $D_E$ denotes a dispersion of the
energy operator $E$. Using Eqs. (\ref{S8})--(\ref{S9}), one obtains
for dynamics of the quantity $\bar{\Gamma}$ in the adiabatic regime,
that \be\label{S10} \frac{d\bar{\Gamma}}{dt}=-D_{\Gamma}^2 \left [
1- \frac{\left ( \overline{E\Gamma}-\bar E\bar \Gamma\right
)^2}{D_E^2D_{\Gamma}^2}\right ].\ee Our goal is to show, that the
function $\bar{\Gamma}(t)$ is non-increasing again, so the system
under consideration has a well-defined arrow of time in the
adiabatic regime of evolution too. To do it, let us show that the
expression $[...]$ in Eq. (\ref{S10}) is not negative. Let us
introduce the formal vector quantities $X$ and $Y$ with the
components $X_n=E_n-\bar E$ and $Y_n=\Gamma_n-\bar \Gamma$, and the
scalar product $(XY)=\sum_nP_nX_nY_n$ of these vectors. It is not
difficult to prove, that in terms of these quantities, the
expression under consideration has the following form:
$[...]=1-(XY)^2/[(XX)(YY)]$. Thus, it is actually non-negative -- in
view of general Cauchy-Buniakowski inequality, which can be applied
to its estimation.

\section{Conclusion}

Thus, in the isothermal and adiabatic regimes of the thermodynamical
evolution, the generalized quantum dynamics developed in this work
has a well-defined arrow of time. Note, that these regimes are the
most important ones for the particle physics and cosmological
applications of the theory \cite{PartPhysCosm-f}. In the particle
physics case, one can take the decay operator in its parity form.
Then, one will deal with the system with pure dynamical mechanism of
the left-right asymmetry production, which seems much more natural
than the standard parity violation scheme. In the cosmological case,
the real presence of the arrow of time in the theory seems necessary
`up to definition'.\, However, the standard cosmology derived from
the General Relativity (or from its supergravity and superstring
modifications) is described by time-reversible equations of motion.
It is clear, that one needs in fundamental generalization of the
theoretical scheme to achieve consistence of cosmology with the
second law of thermodynamics. In this paper, we have proposed the
`minimal quantum theory framework'\, for the  such modification.

We think, that the approach developed above must be applied for
corresponding reformulation of the string theory. In addition to the
`renormalization arguments'\, presented in the section 4 (see Eq.
(4.4)), also one has `pure thermodynamical reasons'\, for the such
string theory generalization. Actually, in string gravity models,
black hole solutions admit the standard thermodynamical
interpretation, related to the Hawking's correspondence between
horizon area and entropy of the black hole. Moreover, in the string
theory this correspondence obtains its fundamental statistical
ground: one can calculate all necessary macroscopic quantities by
counting of their microscopic realizations. Thus, the string theory
itself provides necessary `kinematical base'\, for the conventional
thermodynamical evolution. However, string theory equations of
motion are time-reversible, so its `thermodynamical kinematics'\,
does not support by `thermodynamical dynamics'.\, We think, that the
string theory modification in the direction developed in this paper
can improve this non-natural situation. The resulting stringy
cosmology and black hole physics will in complete agreement with the
standard thermodynamics with the arrow of time.

At the end of this paper we would like to stress, that in the
modified quantum theory under consideration, the symmetry operator
$\Gamma$ (it is taken as commuting with the Hamiltonian of the
system) does not define any conserved quantity. This interesting
property of the quantum theory also must take place in its classical
limit (we hope to demonstrate it in the forthcoming publication).
Thus, our generalization of the theory in the direction of its
irreversibility destroys the well known relation between symmetries
and integrals of motion, which is stated by the famous Noether's
theorem.

\vskip 10mm \noindent {\large \bf Acknowledgements}

\vskip 3mm \noindent e would like to thank prof. B.S. Ishkhanov for
many discussions and private talks which were really useful for us
during this work preparation. One of the authors (O.V.K) was
supported by grant ${\rm MD \,\, 3623.\, 2006.\, 2}$.


\begin{thebibliography}{15}

\bibitem{atf}
A. Magnon, Arrow of time and reality: in search of a conciliation,
Singapore, Singapore: World Scientific (1997).


\bibitem{atf'}
L. Diosi, Lect. Notes Phys. {\bf 633} (2004) 125.


\bibitem{atf''}
M. Castagnino, L. Lara, O. Lombardi, Int. J. Theor. Phys. {\bf 42}
(2003) 2487.



\bibitem{atl}
I. Prigogine, The arrow of time, Published in *Rome/Pescara (1999),
The chaotic universe* 1.


\bibitem{td}
H.B. Callen, Thermodynamics and an introduction to thermodynamics,
2nd edition, John Wiley and Sons, New York (1985).


\bibitem{qm}
Sudbery, E., Quantum mechanics and particles of Nature, Cambridge
Univ. Press, (1986).


\bibitem{cm}
Goldstein, H. Classical mechanics. AddisonWesley Series in Physics,
(1980).


\bibitem{Pr}
I. Prigogine, From Being to Becoming: Time and Complexity in the
Physical Sciences, W. H. Freeman and Co., San Francisco (1980).


\bibitem{ctf}
A. Mejias, L. Di G. Sigalotti, E. Sira, Chaos Solitons Fractals {\bf
19} (2003) 773.


\bibitem{ct-f'}
H. Aoyama, Nucl. Phys. {\bf B446} (1995) 315.

\bibitem{ct-f''}
H. Aoyama, T. Harano, Mod. Phys. Lett. {\bf A10} (1995) 1135.
(1995).


\bibitem{ctl}
S. Biswas, B. Modak, A. Shaw, Gen.Rel.Grav. {\bf 32} (2000) 53.




\bibitem{NonHerm-f}
R. Okamoto, S. Fujii, K. Suzuki,
Int. J. Mod. Phys. {\bf E14} (2005) 21.

\bibitem{NonHerm-f'}
D.B. Fairlie, J. Nuyts,
J. Phys. {\bf A38} (2005) 3611.


\bibitem{NonHerm-f''}
J. Beckers, J. F., N. Debergh, Giuseppe Marmo,\,
Mod. Phys. Lett. {\bf A16} (2001) 91.


\bibitem{NonHerm-l}
C. M. Bender, S. Boettcher,\,
Phys. Rev. Lett. {\bf 80} (1998) 5243.


\bibitem{Str1}
O. V. Kechkin, Phys. Part. Nucl. {\bf 35} (2004) 383.

\bibitem{Str2}
O. V. Kechkin, Phys. Rev. {\bf D65} (2002) 066006.



\bibitem{PartPhysCosm-f}
A. Linde, Particle physics and inflationary cosmology, Chur,
Switzerland: Harwood (1990) Contemporary concepts in physics.

\bibitem{PartPhysCosm-l}
S. W. Hawking, Phys. Rev. {\bf 32} (1985) 2489.

\bibitem{BlH-f}
V. P. Frolov, I. D. Novikov, Black hole physics: basic concepts and
new developments, Kluwer Academic (1998).

\bibitem{BlH-l}
R. Brout, S. Massar, R. Parentani, Phys. Rept. {\bf 260} (1995) 329.

\bibitem{Str-f}
J. R. David, G. Mandal, S. R. Wadia, Phys. Rept. {\bf 369} (2002)
549.


\end{thebibliography}
\end{document}